\documentclass[conference]{IEEEtran}
\IEEEoverridecommandlockouts
\usepackage{tikz}
\usetikzlibrary{positioning}
\usepackage{cite}
\usepackage{amsmath,amssymb,amsfonts}
\usepackage{algorithmic}
\usepackage{graphicx}
\usepackage{textcomp}
\usepackage{xcolor}
\usepackage{tabularx} 
\usepackage{pifont}
\newcommand{\cmark}{\ding{51}}
\newcommand{\xmark}{\ding{55}}
\usepackage{booktabs}
\usepackage[table]{xcolor}
\usepackage{makecell}
\usepackage{enumitem}
\usepackage{comment}
\usepackage[most]{tcolorbox}
\newcommand{\Design}[0]{\textsc{CyberMaskQA}}
\def\BibTeX{{\rm B\kern-.05em{\sc i\kern-.025em b}\kern-.08em
    T\kern-.1667em\lower.7ex\hbox{E}\kern-.125emX}}

\begin{document}

\title{CyberMaskQA: A Privacy-Aware Benchmark for Evaluating Large Language Models in Cybersecurity Question Answering}


\author{\IEEEauthorblockN{Matilda Gaddi, Jin Noh, Onat Gungor, Tajana Rosing}
\IEEEauthorblockA{\textit{Department of Computer Science and Engineering} \\
\textit{University of California, San Diego (UCSD)} \\
\{mgaddi, j4noh, ogungor, tajana\}@ucsd.edu}
}

\maketitle

\begin{abstract}
Large language models (LLMs) are increasingly applied to cybersecurity question answering (QA) for critical tasks such as incident response and vulnerability analysis. However, real-world operational contexts, including system logs and network configurations, inherently contain sensitive identifiers, e.g., IP addresses, host names, and user accounts. Processing this data with cloud-based models is often unsafe or infeasible in regulated environments. Furthermore, progress in privacy-preserving QA is hindered by the lack of annotated, context-rich datasets capable of jointly evaluating operational reasoning and privacy preservation. To address this gap, we introduce \Design{}, a privacy-aware QA benchmark covering key security domains. Unlike existing benchmarks that primarily test factual knowledge, \Design{} grounds questions in realistic organizational contexts with explicit causal dependencies among assets and privileges. Generated through a systematic pipeline, the dataset combines human-curated base scenarios with LLM-driven semantic expansion, annotating each instance with precise private entity labels to enable controlled information disclosure. Evaluations of QA accuracy and masking performance demonstrate the benchmark's utility for developing deployable, context-aware cybersecurity models and facilitating nuanced studies of privacy-utility trade-offs. Upon acceptance, we will release the dataset and the generation framework.
\end{abstract}


\section{Introduction}\label{sec:intro}
Cybersecurity question-answering (QA) refers to the automated process of providing answers to security-related queries, facilitating the rapid identification of potential threats and appropriate remediation strategies \cite{agrawal2024cyberq}. Large language models (LLMs) are increasingly applied to this task due to their ability to reason over complex information \cite{chen2026can,gungor2025aqua,gungor2025eager}. However, real-world operational data, including system logs and incident reports, often contains sensitive identifiers such as IP addresses and user accounts \cite{aghili2024empirical}. Processing such data using cloud-based LLMs can be unsafe or infeasible in operational and regulated environments, creating a pressing need for privacy-preserving QA solutions. Effective cybersecurity QA therefore requires models capable of integrating localized organizational context while strictly protecting sensitive operational data \cite{vielberth2020security}.

As illustrated in Table~\ref{tab:private_qa}, critical tasks such as incident severity assessment and alert prioritization rely on infrastructure attributes that are specific to individual enterprises and are therefore absent from public cybersecurity QA datasets \cite{roumani2025examining}. Because these operational decisions depend on complex relationships among assets, privileges, and network topologies, models trained solely on publicly available datasets tend to overfit to context-agnostic heuristics that fail to generalize to real-world deployments. This discrepancy between benchmark assumptions and operational reality highlights a key limitation of current LLM-based approaches: the lack of access to the structured, organization-specific context required for reliable reasoning. This gap motivates the need for a privacy-preserving QA benchmark that retains the causal dependencies of enterprise environments while anonymizing sensitive information, enabling the development and evaluation of models capable of context-aware cybersecurity reasoning.

\begin{table}[t]
\centering
\caption{Representative operational cybersecurity QA scenarios.}
\label{tab:private_qa}
\small
\renewcommand{\arraystretch}{1.2} 

\begin{tabularx}{.95\columnwidth}{@{}X@{}}
\toprule

\rowcolor{lightgray}
\textbf{Incident Severity} \\
\textit{Q:} A lateral movement attempt is detected between two internal hosts. Does this constitute a high-severity incident? \\
\rowcolor{lightgray}
\textit{Required Context:} Asset roles, network segmentation, user privilege levels. \\
\midrule

\rowcolor{lightgray}
\textbf{Alert Prioritization} \\
\textit{Q:} Malware is detected on an executive device while a login anomaly occurs on a gateway system. Which alert should be prioritized? \\
\rowcolor{lightgray}
\textit{Required Context:} Asset criticality, system dependencies, operational impact. \\

\bottomrule
\end{tabularx}
\end{table}

Most existing cybersecurity QA benchmarks~\cite{liu2023secqa,tihanyi2024cybermetric,alam2024ctibench,liu2024cyberbench} are constructed from textbooks, standards, and publicly available documentation. Consequently, they primarily evaluate factual knowledge (e.g., ``Which port does HTTPS use?'') rather than operational reasoning. While useful for assessing conceptual understanding, these benchmarks fail to capture the handling of sensitive, localized information required in realistic deployment settings. Recent efforts, including PRIV-QA~\cite{li2025privqa} and CON-QA~\cite{singh2025conqa}, propose frameworks incorporating privacy-preserving mechanisms. However, these approaches rely on disjoint or narrowly scoped datasets and do not adequately address scenarios where contextual reasoning and privacy preservation must be jointly enforced. This limitation hinders the evaluation of QA systems that are both practically deployable and robust in real-world cybersecurity applications.

To address these limitations, we propose the \Design{} framework (Figure~\ref{fig:framework}), a systematic pipeline for constructing privacy-aware cybersecurity QA benchmarks. The framework standardizes dataset structure via schema definition and collects domain-specific operational documentation to ground the language. We then design base questions paired with explicit organizational context, where accurate answers depend directly on that context. To ensure diversity and scale, we algorithmically expand these base questions into multiple semantically consistent variations. Finally, we generate privacy-preserving, masked versions of each instance. This pipeline yields the initial \Design{}-250 dataset, which we leverage alongside iterative validation stages to scale and construct the comprehensive \Design{}-2000.

The \Design{} benchmark supports multiple evaluation paradigms, addressing key limitations of prior work and enabling advances in privacy-preserving cybersecurity QA: (i) it enables context-aware evaluation through multiple-choice questions grounded in realistic organizational settings; (ii) it provides labeled sensitive information annotations to facilitate the evaluation of privacy-preserving masking techniques; and (iii) it supports integrated evaluation across tasks. Furthermore, our initial empirical evaluation using \Design{} exposes a critical vulnerability in current edge-deployed LLMs: a consistent trade-off where reasoning accuracy comes at the direct expense of privacy preservation.
\section{Related Work}\label{sec:background}

\subsection{Cybersecurity QA Datasets}
\textbf{CyberMetric}~\cite{tihanyi2024cybermetric} is a collection of multiple-choice QA datasets ranging from 80 to 10K questions. Questions were generated using GPT-3.5 with Retrieval-Augmented Generation (RAG), drawing from credible sources including NIST standards and textbooks. CyberMetric primarily evaluates factual knowledge and does not incorporate institutional context or privacy-sensitive information. 

\textbf{CyberBench}~\cite{liu2024cyberbench} comprises ten datasets across four tasks: named entity recognition, summarization, multiple-choice QA, and text classification. While it provides task diversity, each dataset is separated by task, preventing integrated evaluation of multi-task frameworks. Furthermore, CyberBench does not support context-aware reasoning or privacy evaluation.

\textbf{SensitiveQA}~\cite{li2025privqa}, part of PRIV-QA, contains 50k Chinese and 7k English questions, including personally identifiable information (PII). The dataset uses background texts and consists of open-ended dialogue-style entries rather than multiple-choice questions, requiring complex semantic evaluation pipelines. PII is present but not labeled, and the dataset is closed source, limiting reproducibility.

\textbf{AttackQA}~\cite{krishna2024attackqa} provides 25K QA pairs with supporting rationales, focused on cybersecurity and physical attacks. About 80\% of the dataset was synthetically generated using Llama3-8B, grounded in the MITRE ATT\&CK knowledge base, with quality control performed via a larger LLM. However, the dataset lacks masking annotations for sensitive information.

\subsection{Limitations of Existing Benchmarks}
Although prior datasets support cybersecurity QA or privacy-sensitive text, they do not enable joint evaluation of context-aware security reasoning and privacy preservation. As summarized in Table~\ref{tab:dataset_comparison}, most existing datasets either: 
(i) focus on factual knowledge without institutional context (e.g., CyberMetric, CyberBench), 
(ii) include sensitive information without labeled masking (e.g., SensitiveQA), or 
(iii) are not openly accessible, limiting reproducibility. The table highlights that none of the prior datasets combine cybersecurity QA, operational context, and explicit privacy annotations in a unified framework. This gap underscores the need for a dataset which enables controlled measurement of privacy--utility trade-offs and supports realistic evaluation of LLMs.

We introduce \Design{}, a privacy-preserving cybersecurity QA dataset that differs from existing benchmarks by combining context-aware reasoning with explicit masking. \Design{}-250 consists of 250 multiple-choice questions requiring institutional context, with both PII and operational information annotated. The dataset is constructed using a hybrid human-synthetic approach and expanded through validated variations to ensure diversity. Masked and labeled versions enable fine-grained evaluation of privacy--utility trade-offs. The framework further scales to \Design{}-2000, supporting studies of context-aware QA and privacy-preserving strategies in realistic scenarios.

\begin{table}[t]
    \centering
    \caption{Comparison between \Design{} and related datasets.}
    \label{tab:dataset_comparison}
    \scriptsize
    \setlength{\tabcolsep}{2.5pt}
    \resizebox{\columnwidth}{!}{
    \begin{tabular}{l c c c c c c c c}
        \toprule
        Dataset & CyberSec & QA & \thead{Context /\\ Realism} & \thead{Private Data\\ Detection /\\ Masking} 
        & Synthetic & \thead{Open\\ Source} & \thead{Human\\ Verified} & \thead{Total} \\
        \midrule
        SecQA \cite{liu2023secqa} & \cmark & \cmark & \xmark & \xmark & \cmark & \cmark & ? & 242 \\
        CyberMetric \cite{tihanyi2024cybermetric} & \cmark & \cmark & \xmark & \xmark & \cmark & \cmark & 500 & 10000 \\
        AttackQA \cite{krishna2024attackqa} & \cmark & \cmark & ? & \xmark & \cmark & \cmark & 5071 & 25355 \\
        CyberBench \cite{liu2024cyberbench} & \cmark & \cmark & \xmark & \cmark & \cmark & \cmark & ? & 244 \\
        CyberLLaMa \cite{zhang2025cyberllama} & \cmark & \xmark & \xmark & \cmark & \cmark & \xmark & ? & 42404 \\
        SensitiveQA \cite{li2025privqa} & \xmark & \xmark & ? & \cmark & \cmark & \xmark & ? & 7214 \\
        pii-masking-300k \cite{ai4priv2024piimask300} & \xmark & \xmark & \cmark & \cmark & ? & \cmark & ? & 300000\\
        CyberMaskQA [ours] & \cmark & \cmark & \cmark & \cmark & \cmark & \cmark & 250 & 2000 \\
        \bottomrule
    \end{tabular}
    }
\end{table}
\begin{figure*}[tp]
\centering
 \includegraphics[width=.83\linewidth]{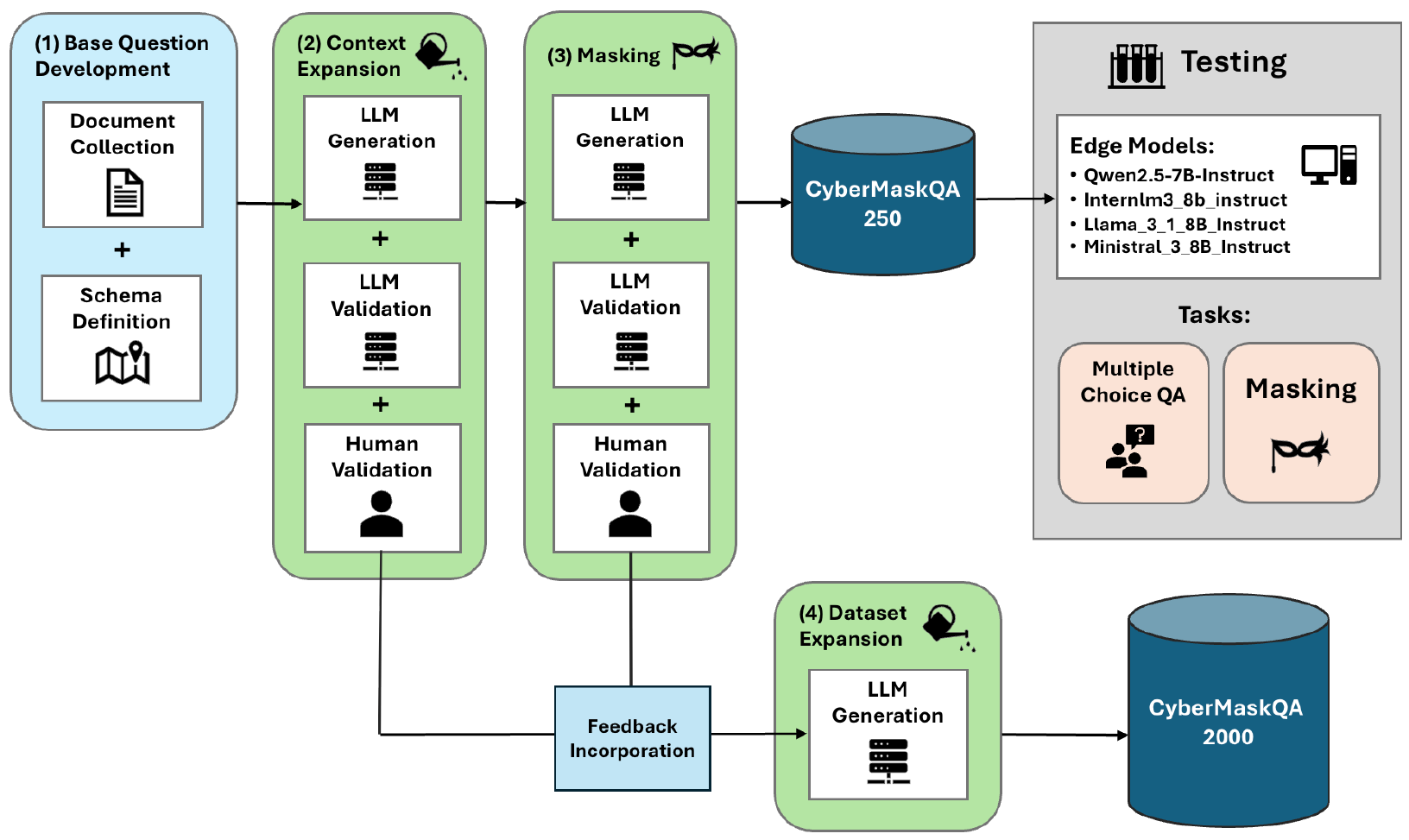}
 \caption{Overview of \Design{} generation framework.}
\label{fig:framework}
\vspace{-10pt}
\end{figure*}

\section{\Design{} Architecture}\label{sec:dataset}

Figure~\ref{fig:framework} presents the \Design{} framework, structured into four stages:
\begin{enumerate}[noitemsep,nolistsep]
    \item \textbf{Base Question Development:} This stage establishes the foundation of the dataset by curating an initial set of questions for subsequent expansion. The questions are organized into four domains: Identity \& Access Management, Threat Detection \& Incident Analysis, Incident Response \& Escalation, and Security Posture \& Prevention. These base questions are grounded in operational standards such as HIPAA, OWASP, and NIST.
    \item \textbf{Context Expansion:} In this stage, advanced LLMs are used to augment the initial questions with realistic scenarios that incorporate enterprise settings and private data. The generated data is subsequently validated by human annotators to ensure quality and consistency.
    \item \textbf{Masking:} This stage applies a masking protocol to anonymize sensitive information. Identifiers such as usernames and IP addresses are replaced with standardized tags to preserve privacy. 
    \item \textbf{Dataset Expansion:} Finally, the dataset is expanded by incorporating feedback from the human validation process, enhancing both coverage and realism. This stage scales the initial \Design{}-250 dataset into the comprehensive \Design{}-2000.
\end{enumerate}

\subsection{Stage 1: Base Question Development}

\subsubsection{Document Collection} We begin by collecting and analyzing official cybersecurity and privacy documentation to ground question development in real-world practices. To ensure technical accuracy and regulatory relevance, dataset scenarios are aligned with authoritative standards, security frameworks, and operational procedures. This grounding ensures that evaluation reflects the complexity of real-world security decision-making and compliance requirements. The primary sources used in this process include:

\begin{itemize}[noitemsep,nolistsep]
    \item \textbf{NIST Special Publications:} Security configurations and control requirements are derived from NIST guidelines (e.g., SP~800-53), reflecting widely adopted industry security practices.
    
    \item \textbf{CIS Critical Security Controls:} Baseline operational security controls are incorporated to model practical defensive measures.
    
    \item \textbf{OWASP Resources:} Vulnerability patterns and log-based attack scenarios are informed by OWASP materials (e.g., OWASP Top~10) and related records, to ensure realistic and representative web security threats.
    
    \item \textbf{ISO/IEC Standards:} International best practices for information security management are incorporated through ISO/IEC standards.
    
    \item \textbf{GDPR:} Data protection and breach notification scenarios are aligned with relevant GDPR provisions, such as Article~33, which defines reporting requirements following a data breach.
    
    \item \textbf{HIPAA:} Healthcare-specific regulatory requirements, including breach notification procedures, are modeled based on HIPAA guidelines.
    
    \item \textbf{Institutional Security Practices:} Additional scenarios are informed by anonymized operational procedures observed within a university information technology (IT) environment to capture realistic administrative workflows.
\end{itemize}

\subsubsection{Schema Definition} The framework defines a structured schema to standardize all dataset entries across cybersecurity domains. The schema specifies mandatory fields for a consistent format for utility and evaluation. Each entry includes:

\begin{itemize}[noitemsep,nolistsep]
    \item \textbf{Context}: Structured system logs (e.g., CSV format)
    \item \textbf{Question}: Context-dependent query
    \item \textbf{Answer Options}: Multiple-choice options (A--D)
    \item \textbf{Correct Answer}: Ground-truth label
    \item \textbf{Entities}: Annotated named entities within the context
    \item \textbf{Masked Fields}: Masked version of the context, question, and answer
\end{itemize}

\noindent An example of the operational schema is shown in Figure~\ref{fig:operational_schema_example}. 

\begin{figure}[t]
\centering
\noindent\fbox{%
\begin{minipage}{0.95\columnwidth}
\small
\raggedright
\textbf{Context (Log):} A structured log (\textit{VPNAccessLog.csv}) identifies user \textit{k\_patel} using an \textit{unmanaged}, \textit{personal Android Pixel} device.

\vspace{0.5em}
\textbf{Context (Policy):} An unstructured \textit{Intranet Access Policy} restricts internal portals to \textit{managed} devices only.

\vspace{0.5em}
\textbf{Question:} Should the access request from \textit{k\_patel} be permitted based on the provided logs and policy?

\vspace{0.5em}
\textbf{Answer Options:} 
(A) Yes, if the user uses a VPN; 
(B) No, the device status is unmanaged; 
(C) Yes, members of the IT unit have unrestricted access; 
(D) No, the user role is unauthorized.

\vspace{0.5em}
\textbf{Correct Answer:} (B)

\vspace{0.5em}
\textbf{Entities:} 
\texttt{\{[USER\_ACCOUNT]: k\_patel, [HARDWARE]: Android Pixel\}}

\vspace{0.5em}
\textbf{Logic:} The model must parse the \textit{unmanaged} device status from the log and apply the \textit{Intranet Access Policy}, rather than relying on context-agnostic heuristics.
\end{minipage}%
}
\caption{Operational Schema Example}
\label{fig:operational_schema_example}
\end{figure}

\noindent\textbf{Domain Categories.}
The questions are organized into four domains. This organization aligns with the workflow of security operations center analysts and ensures that models are evaluated on tasks requiring context-aware reasoning and realistic operational decision-making.
\begin{enumerate}[noitemsep,nolistsep]
    \item \textbf{Identity \& Access Management}: This domain covers the lifecycles of user permissions, access control lists based on role, and enforcing the Principle of Least Privilege. Including this domain ensures that the model is tested on the ability to determine if a specific user possesses the privilege or authorization to perform a requested action. \Design{}-250 contains 76 entries of this category and 840 labeled private entities.
    \item \textbf{Threat Detection \& Incident Analysis}: Questions in this category focus on evaluating whether the model is able to recognize suspicious patterns within technical logs. Examples include being able to identify SQL injection and Cross Site Scripting (XSS) attacks. \Design{}-250 includes 48 entries of this category and 418 labeled private entities.
    \item \textbf{Incident Response \& Escalation}: This domain evaluates compliance with regulatory standards regarding reporting timelines and organizational protocols. It combines standards from organizations such as HIPAA and GDPR to reason on the severity of an event and determine the next steps to remain compliant with industry standards. \Design{}-250 consists of 47 entries of this category and 282 labeled private entities.
    \item \textbf{Security Posture \& Prevention}: This category assesses hardening strategies to increase organizational security. It includes evaluating network architectures, mobile device management policies, and prioritization for security focused training. \Design{}-250 contains 79 entries of this category and 601 labeled private entities.
\end{enumerate}
The difference in the number of entries per category is due to the variance of realistic scenarios for each domain. 





Our framework establishes operational relevance by constructing 30 base scenarios modeled on real-world enterprise security environments. Operational relevance is supported in the data set by having organizational attributes as context. Organizational attributes in the dataset are defined as data segments containing structured system logs through a CSV file, unstructured organizational policies, and a context-dependent query. Figure~\ref{fig:organizational_attributes_example} illustrates this type of context-dependent reasoning. 
From the system log, policy reference, and deletion request, the model must recognize that the data corresponds to a medical record and consult the relevant HIPAA retention requirements before deciding whether the file can be deleted.

\begin{figure}[t]
\centering
\fbox{%
\begin{minipage}{0.95\columnwidth}
\small
\raggedright
\textbf{System Logs:} \textit{patientRecord.csv}

\vspace{0.4em}
\textbf{Policy:} \textit{Medical Records Policy}

\vspace{0.4em}
\textbf{Context-dependent Query:} ``We have received a request to delete personal data for user PT-2156. Should we honor this request?''
\end{minipage}%
}
\caption{Organizational Attributes Example}
\label{fig:organizational_attributes_example}
\end{figure}

\subsection{Stage 2: Context Expansion}

\subsubsection{LLM Generation} The second stage of our framework scales the foundational set of base questions into a diverse set of operational scenarios. The objective of the expansion is to introduce different contexts while preserving the logical dependencies of the foundational scenarios. We use \texttt{Claude Sonnet 4.5} for this stage for its technical reasoning and ability to consistently generate the dataset schema and structure correctly. 
By generating multiple variations from a single foundational question, the framework enables evaluation of models across diverse contexts, emphasizing reasoning capabilities rather than reliance on memorized heuristics. An example illustrating multiple contextual variations derived from the same base question is provided in Figure~\ref{fig:context_difference_example}. 

\begin{figure}[t]
\centering
\fbox{%
\begin{minipage}{0.95\columnwidth}
\small
\raggedright

\textbf{Version 1: Employee Record}

\textbf{Question:} ``We have received a request to delete personal data for user EMP-3301. Should we honor this request?''

\vspace{0.3em}
\textbf{Context:} The file \textit{EmployeeRecords.csv} lists EMP-3301 as an active employee in the Engineering department. The \textit{Employee Data Policy} states that employee data cannot be deleted while the employee is actively employed.

\vspace{0.3em}
\textbf{Correct Answer:} No. The user is an active employee, so the deletion request cannot be honored.

\vspace{0.8em}
\textbf{Version 2: Customer Record}

\textbf{Question:} ``We have received a request to delete personal data for user CUST-4521. Should we honor this request?''

\vspace{0.3em}
\textbf{Context:} The file \textit{CustomerData.csv} lists CUST-4521 as an inactive customer located in Paris. The \textit{GDPR Compliance Policy} states that deletion requests from inactive EU customer accounts can be processed upon request.

\vspace{0.3em}
\textbf{Correct Answer:} Yes. The user is an inactive EU customer, so the deletion request should be honored.
\end{minipage}%
}
\caption{Context Difference Example.}
\label{fig:context_difference_example}
\end{figure}


\subsubsection{LLM Validation}  
Prior to expansion to the full dataset, we assess the robustness of LLM validation and detect potential agreement bias by creating small, controlled perturbations, intentionally modifying selected questions, answers, and explanations to create incorrect instances. We test this approach on two models. \texttt{GPT-5.2 Instant} missed 3 of the 10 perturbations while \texttt{Gemini 3 Flash} consistently identified every perturbed entry as incorrect. Hence, we utilize \texttt{Gemini 3 Flash} to review the full set of generated entries. Results from Gemini validation are used to flag potentially problematic entries in the dataset to improve the speed of human validation and correction.

\subsubsection{Human Validation} After reviewing the questions flagged by Gemini, human validators assess all of the generated questions by hand. Problematic questions were corrected, and ten questions were removed from the 260 previously generated. This stage serves as a manual validation step to identify and correct errors that may persist after LLM-based validation. The human validation process focuses on three primary aspects:

\begin{enumerate}[noitemsep,nolistsep]
    \item \textbf{Policy Grounding:} Each scenario referencing authoritative standards from Stage 1 is verified against the original standard to ensure accurate interpretation. Human validators confirm that the correct answer is the only logically valid choice based on both the standard and the scenario context. Examples of regulatory frameworks include HIPAA and NIST.
    
    \item \textbf{Uniqueness:} To prevent models from learning repetitive patterns, human validators refine or remove scenarios that are overly similar. Refinement may include modifying attributes such as incident timestamps or changing accessed assets, e.g., different software or systems.
    
    \item \textbf{Operational Fidelity:} Human validators ensure that identifiers, such as IP addresses and usernames, are realistic and contextually appropriate.
\end{enumerate}

\textbf{Context Correction.} The validation process also identifies common errors to improve subsequent dataset expansions. Addressing these errors ensures technical integrity and enhances accuracy in the final benchmark. Discrepancies include unrealistic personally identifiable information (PII), such as placeholder usernames or IP addresses that do not match the geographic location in the scenario. Other issues involve multiple-choice options: some entries contain ambiguous logic with multiple plausible answers, while others exhibit linguistic bias, where the correct option is noticeably longer than the distractors. Figure~\ref{fig:context_correction_example} presents an example illustrating how the context was corrected.

\begin{figure}[t]
\centering
\fbox{%
\begin{minipage}{0.95\columnwidth}
\footnotesize
\raggedright

\textbf{Incorrect Context:}

\vspace{0.2em}
\texttt{UserSessions.csv: Username, Role, IPAddress1, IPAddress2, Location1, Location2, Status}

\texttt{contractor\_k\_patel, Contractor, 185.92.44.12, 185.92.44.13, Berlin, Germany, Berlin, Germany, Active}

\vspace{0.5em}
\textbf{Corrected Context:}

\vspace{0.2em}
\texttt{UserSessions.csv: Username, Role, IPAddress1, IPAddress2, Location1, Location2, Status}

\texttt{contractor\_k\_patel, Contractor, 31.16.188.104, 31.16.188.108, Berlin, Germany, Berlin, Germany, Active}

\vspace{0.5em}
\textbf{Logic:} The scenario specifies Berlin, Germany as the user location, but the original IP addresses were inconsistent with this geographic context. We therefore replace them with IP addresses that better align with the intended location, improving the realism and internal consistency of the generated scenario.
\end{minipage}%
}
\caption{Context Correction Example}
\label{fig:context_correction_example}
\end{figure}

    
    


\subsection{Stage 3: Masking}

The masking stage of the data generation process is designed to align with \Design{}'s objective of evaluating a model's ability to preserve privacy. Validated entries are transformed into a sanitized format in which sensitive information is replaced with standardized masking labels. This process targets cybersecurity-specific identifiers, such as IP addresses and usernames, replacing them with tokens.





The framework supports a comprehensive set of sensitive identifiers, each mapped to operational requirements across the four domains defined in Stage 1
. Table~\ref{tab:masking_labels} lists all masking labels and their associated descriptions. By standardizing these labels, the framework ensures privacy-preserving data handling while maintaining the contextual integrity required for analysis and decision-making. The masked components of the dataset are designed to be customizable, allowing users to specify which entity types to mask. For instance, if \texttt{[GROUP]} entities are considered low priority for privacy, they can be excluded from evaluation. Users may choose fully masked, unmasked, or develop hybrid versions of the dataset depending on their privacy requirements. 

\begin{table}[t]
\centering
\caption{Masking labels and their corresponding sensitive data types.}
\label{tab:masking_labels}
\small
\setlength{\tabcolsep}{4pt}
\begin{tabular}{p{0.28\columnwidth} p{0.62\columnwidth}}
\toprule
\textbf{Label} & \textbf{Description} \\
\midrule
USER\_ACCOUNT & Person names, employee IDs, usernames, user IDs \\
IP\_ADDRESS & IP addresses, optionally including ports \\
URL & Web addresses \\
FILE\_PATH & File or directory paths \\
HOSTNAME & Network hostnames \\
SOFTWARE & Application or system software names \\
OS & Operating systems \\
HARDWARE & Devices, servers, or computing resources \\
GROUP & Project names, teams, roles \\
IDENTIFIER & Context-specific identifiers, such as incident IDs or device IDs \\
LOCATION & Geographical locations \\
DATE & Timestamps and dates \\
\bottomrule
\end{tabular}
\end{table}

To transform raw cybersecurity scenarios into an anonymized format, masking is applied with \texttt{Claude 4.5 Sonnet}. We provide three human-developed examples of masked dataset items to Claude and prompt it to apply the masking portion of the schema defined in Section 3.1.B, using the entity labels and definitions defined in Table~\ref{tab:masking_labels}. The "Question" field of the dataset is structured as one or two sentences of text, often containing one or more private entities. In this case, the corresponding "Masked Question" field would contain the same exact text, substituting the private entities for their masked label. This process is applied similarly to the "Context" and answer option fields. The masking process maintains the structure of the original data while replacing sensitive entities.


\textbf{Value-to-Label Mapping.} For every dataset entry, the framework maintains a label dictionary that maps original sensitive values (e.g., username: ``k\_patel'') to their corresponding bracketed labels (e.g., \texttt{[USER\_ACCOUNT]}). This lookup table is an important facilitator for assessing a model's data masking performance for privacy preservation for the context, question, and answer fields of the dataset. Once the data has been masked, it undergoes a similar validation process to the Context Expansion stage to ensure correct masking and to prevent leakage of sensitive information.

\textbf{Validation.} Masking is verified using a protocol that combines automated and human checks. Human validators flag and fix issues with masking correctness and logical consistency. This is followed by automated evaluation to catch missed cases with \texttt{Gemini 3 Flash}. Human validators then review flagged entries to detect privacy leaks or mapping errors, ensuring both efficiency and accuracy.

Human validation of 250 dataset entries identified 115 masking failures requiring manual correction. A major source of these errors is a structural phenomenon termed the ``Foundational Carry Over'' effect: when a base question contains a missed sensitive value, the error can propagate across its derivative variations. Correcting the base question quickly alleviates and prevents repeated failures across related variations.

\subsection{Stage 4: Dataset Expansion}

The dataset expands from the initial 250 questions to 2,000 through a structured expansion pipeline to preserve operational context while diversifying scenarios. This stage leverages error patterns identified during previous validation phases to ensure that the expanded dataset maintains operational realism and consistently applies privacy-preserving masking.

To mitigate redundancy inherent in large-scale expansion, the library of foundational questions grows to 60 base questions. Each base question is expanded into 30--40 distinct contexts, exceeding the original 10 contexts used in the 250-question dataset. This additional variation captures differences in LLM performance on similar scenarios, such as divergent question accuracy or masking behavior, thereby enabling more precise evaluation of model capabilities. 

Foundational questions are generated using Retrieval-Augmented Generation (RAG) to ground each question in operationally relevant context. The model ingests records of cybersecurity standards, guidelines, and attack reports from authoritative organizations including OWASP and NIST. It then generates sets of 10 foundational questions based explicitly on the ingested documents. This process continues until 70 new foundational questions are created.

The expansion of these foundational questions follows the same methodology as the original 30-question set, with key modifications informed by prior error patterns. Prompting is adjusted based on human validation feedback to address common issues seen while developing the 250 question dataset, including ensuring IP addresses align with locations, avoiding redundant questions, and eliminating placeholder values. Once the additional questions are generated, the dataset proceeds to the masking phase.

Masking in the expanded dataset parallels the initial masking procedure, with prompt adjustments based on human feedback to address recurrent issues identified during automated masking. These include associating the [GROUP] entity with professional roles, correctly masking member references, and careful handling of the [LOCATION] entity. The final expanded dataset constitutes a comprehensive benchmark that supports rigorous evaluation of both utility and privacy-preserving performance.

\section{Experimental Analysis}\label{sec:results}

\subsection{Experimental Setup}
We evaluate CyberMaskQA-250 using four LLMs: Llama-3.1-8B-Instruct, InternLM3-8B-Instruct, Mistral-3-8B-Instruct, and Qwen2.5-7B-Instruct. These open-source models provide an optimal balance between reasoning capacity and computational efficiency, enabling multi-step reasoning and multi-entity masking while remaining deployable in resource-constrained, edge environments. Focusing on edge-capable models reflects the operational requirements of cybersecurity applications, where local inference preserves privacy, reduces latency, and avoids transmitting sensitive data to external cloud services. This selection ensures that \Design{} is evaluated under realistic operational constraints while still challenging the reasoning and privacy-preserving capabilities.

For MCQA, we instruct models to output only the letter (A, B, C, or D) corresponding to the correct answer, and evaluate performance using \textbf{accuracy.} For masking, models generate JSON-formatted outputs in which the Context, Question, and answer options replace sensitive entities with standardized placeholder tokens, e.g., \texttt{[IP\_ADDRESS]}, consistent with prior datasets \cite{ai4priv2024piimask300}. We employ one-shot prompting for both MCQA and masking, presenting the model with a single example prompt–response pair before providing the new question for evaluation. Following prior work \cite{li2025privqa,zhang2025cyberllama}, we quantify masking performance using \textbf{precision} (correctly masked / all masked) and \textbf{recall} (correctly masked / all ground truth). While high recall indicates comprehensive identification of sensitive entities, high precision reflects minimal false positives, thereby preserving the utility of the anonymized data.

\subsection{Multiple Choice QA Results}

\subsubsection{Overall Performance} Figure~\ref{fig:overall-mcqa} presents MCQA performance across all question categories, along with their average performance. On average, Llama-3.1-8B achieves the highest accuracy (78.0\%), followed by InternLM3-8B (76.4\%) and Qwen2.5-7B (75.2\%). Ministral-3-8B performs substantially lower at 40.8\%, likely due to its training focus on long-form generative outputs rather than constrained single-token classification tasks. Ministral often produces verbose justifications prior to the single-letter answer, leading to extraction failures even when the underlying reasoning is correct.

\subsubsection{Category-Level Analysis} 
\begin{figure}[t]
    \centering
    \includegraphics[width=\columnwidth]{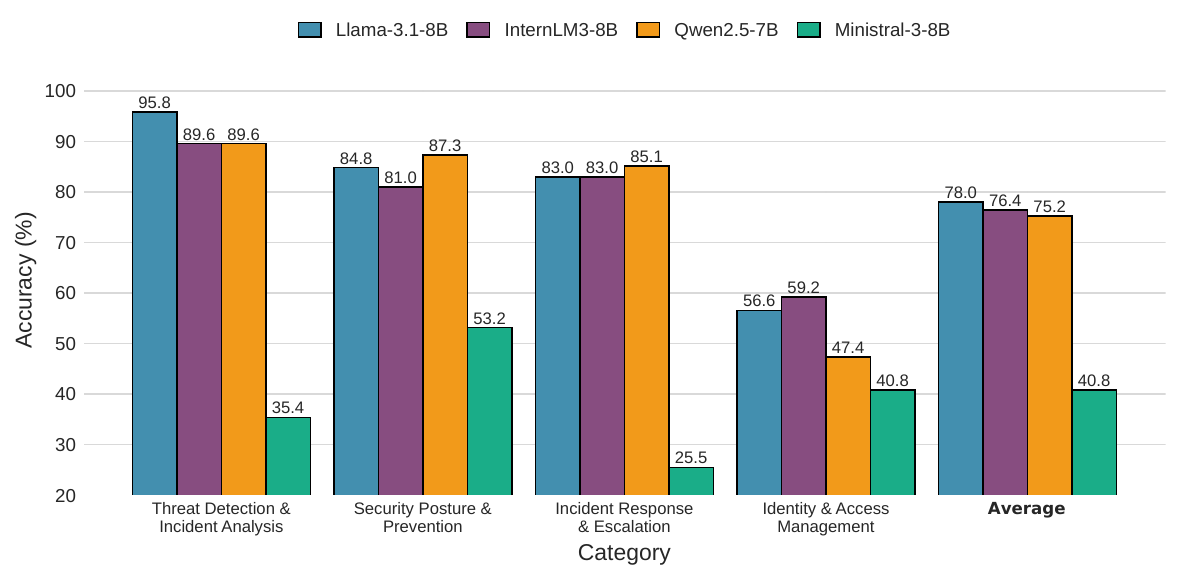}
    \caption{MCQA accuracy across the four \Design{} categories.}
    \label{fig:overall-mcqa}
\end{figure}

Based on Figure~\ref{fig:overall-mcqa}, most models perform best on Threat Detection \& Incident Analysis Llama: 95.8\%, InternLM: 89.6\%, Qwen: 89.6\%), likely due to clearer contextual evidence, which reduces the number of inferences required to answer the questions. Identity \& Access Management is the most challenging category because correct answers require integrating information across multiple documents: a user's group membership in one file, the group's permissions in another, and the resource's requirements in a third. For instance, determining whether user \texttt{tjones} can access a restricted database requires cross-referencing their role assignment, the role-to-permission mapping, and the database's access-control policy. This reasoning pattern is handled inconsistently by the selected LLMs.


\subsection{Masking Results}

\subsubsection{Overall Performance} 
Figure~\ref{fig:overall-masking} illustrates the precision–recall trade-off. Overall, Ministral-3-8B achieves the highest average recall (92.5\%) but exhibits lower precision (61.7\%) due to over-masking. This behavior reflects the model's emphasis on privacy preservation during prompting, prioritizing high recall (minimizing private information leakage) over high precision (reducing unnecessary masking of non-sensitive information).

\subsubsection{Category-Level Analysis} 
\begin{figure}[t]
    \centering
    \includegraphics[width=\columnwidth]{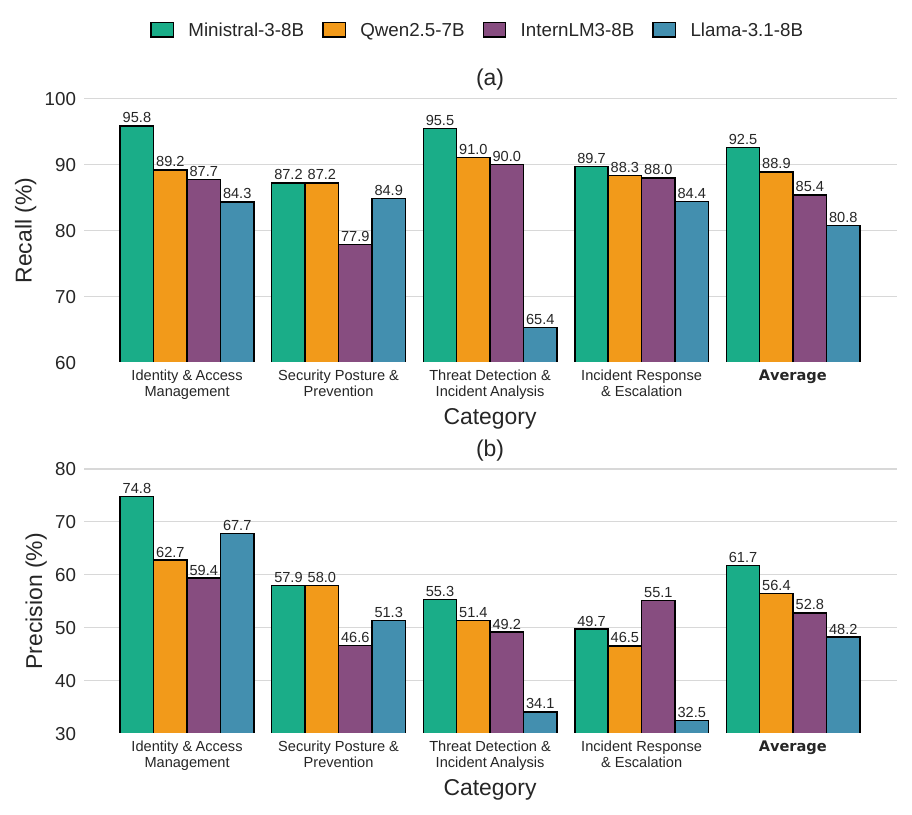}
    \caption{Masking performance across the four \Design{} categories.}
    \label{fig:overall-masking}
\end{figure}

As shown in Figure~\ref{fig:overall-masking}, models consistently perform well on Threat Detection \& Incident Analysis, as questions in this category typically rely on a single, unambiguous contextual signal (e.g., a log entry indicating a failed authentication from an external IP targeting a privileged account), leaving little room for alternative interpretations. Identity \& Access Management yields the most consistent masking results because its entities (usernames, badge IDs, IP addresses) follow rigid, recognizable formats that models can pattern-match reliably. Security Posture \& Prevention is substantially harder because policy documents blend organizational jargon with common English nouns. Terms like ``threat actor,'' ``privileged user,'' or ``critical asset'' are semantically sensitive in context but do not carry the lexical or structural signatures (e.g., fixed length, delimiter-bounded, capitalized prefix) that facilitate identification as a private data point. Models switch between treating them as sensitive entities and treating them as generic technical vocabulary, producing inconsistent precision.

\subsubsection{Entity-Level Analysis}
\begin{figure}[t]
    \centering
    \includegraphics[width=\columnwidth]{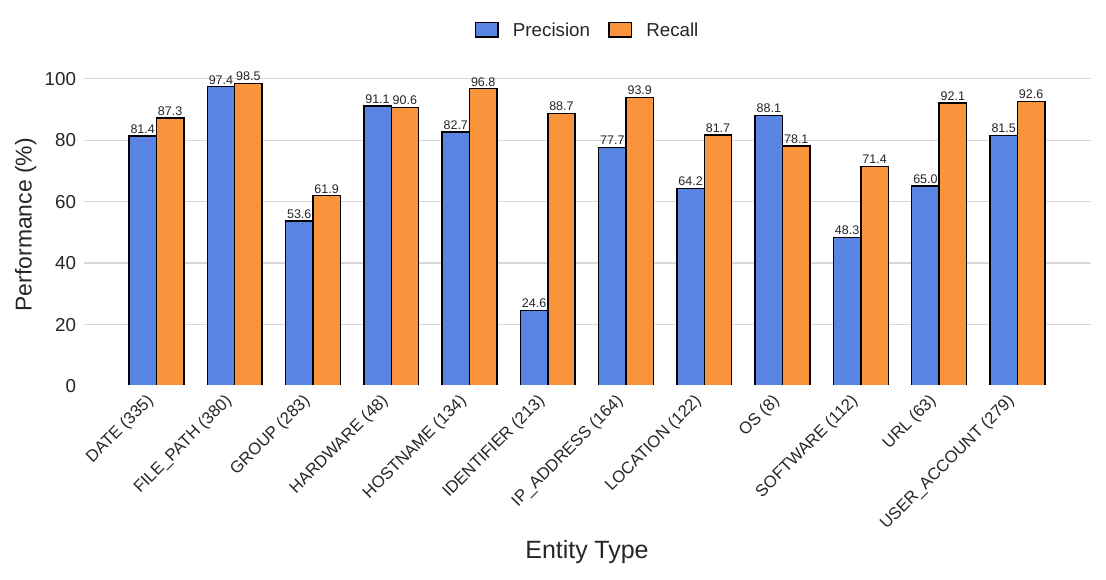}
    \caption{Entity-level masking performance averaged across the four LLMs.}
    \label{fig:by_entity}
\end{figure}

Entity-level analysis, averaging precision and recall across all models (Figure~\ref{fig:by_entity}), reveals substantial variation. Entities with consistent formatting and labeling patterns, such as \texttt{IP\_ADDRESS} and \texttt{FILE\_PATH}, achieve higher masking performance, whereas entities with more varied formats and common nouns, such as \texttt{IDENTIFIER} and \texttt{GROUP}, exhibit lower performance.

IDENTIFIER entities (incident IDs, ticket numbers, equipment IDs such as \texttt{LAB-SEQ-07} or \texttt{INC-20481}) show this ambiguity challenge. Unlike \texttt{IP\_ADDRESS} or \texttt{FILE\_PATH}, which carry structural markers (dotted-quad notation, directory separators), identifiers share no universal format. They may be purely numeric, alphanumeric, or prefixed. Models achieve high recall (79--95\%) by applying a broad heuristic: mask any token resembling an ID. This sweeps in true identifiers but also catches innocent numeric strings. In Q93, for instance, the Llama correctly masks the \texttt{[USER\_ACCOUNT]} field but then over-masks action labels such as \texttt{VPNConnect} and \texttt{LoginSuccess} as if they were sensitive identifiers, while leaving IP addresses and timestamps entirely unmasked, demonstrating that the model has learned a surface-level heuristic rather than a semantic understanding of what constitutes sensitive data. This yields low precision for IDENTIFIER entities, suggesting that reliable masking of this type may require rule-based post-processing rather than relying solely on the LLM judgment.

\subsection{Discussion}

\subsubsection{Inverse Task Performance} 
\begin{figure}[t]
    \centering
    \includegraphics[width=.97\columnwidth]{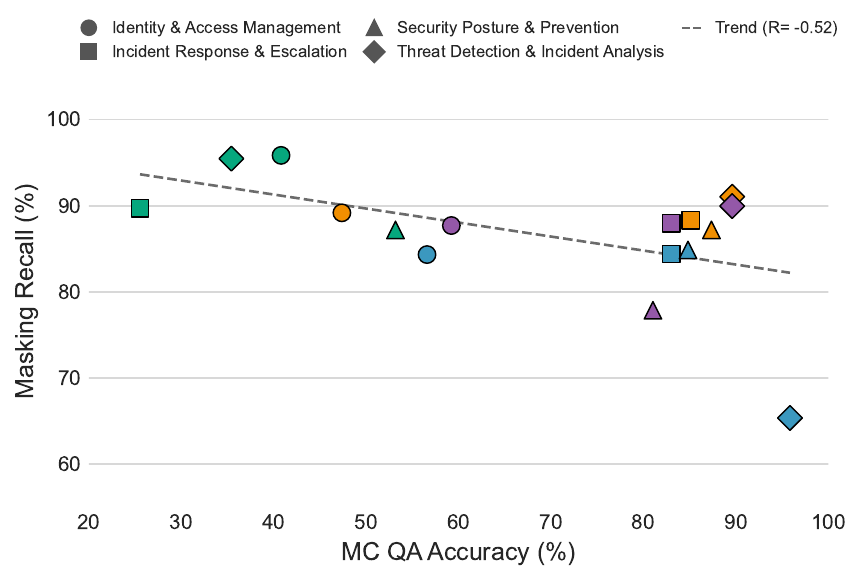}
    \caption{Relationship between QA accuracy and masking recall.}
    \label{fig:acc_recall}
\end{figure}

A notable finding is that the best-performing MCQA model (Llama-3.1-8B) performs worst on masking, whereas the lowest-performing MCQA model (Ministral-3-8B) excels in masking. Figure~\ref{fig:acc_recall} illustrates a negative correlation between MCQA and masking performance across models and categories. This inverse relationship suggests that the two tasks rely on fundamentally different capabilities: QA benefits from precise reasoning, whereas masking favors conservative generation patterns.

\subsubsection{Precision-Recall Tradeoff} Ministral-3-8B and Qwen2.5-7B prioritize privacy, exhibiting high recall but lower precision, whereas Llama-3.1-8B favors precision at the expense of recall. The optimal operating point depends on whether false negatives (private data leakage) or false positives (unusable output) are more costly. Entities such as \texttt{FILE\_PATH} and \texttt{DATE} can be reliably masked with current models (recall $>95\%$), whereas \texttt{IDENTIFIER} entities require rule-based approaches due to low precision.

\section{Conclusion}\label{sec:conclusions}
We present \Design{}, a comprehensive benchmark dataset for evaluating large language models on operational enterprise cybersecurity tasks, specifically context-aware question answering and privacy preservation. Constructed via a systematic, multi-stage pipeline, the scaled \Design{}-2000 dataset comprises operationally realistic questions spanning multiple security domains. Crucially, each instance is explicitly grounded in organizational context and contains precisely annotated sensitive information requiring protection. Our empirical evaluation of four state-of-the-art edge models reveals a consistent trade-off between reasoning performance and privacy preservation: models achieving higher question-answering accuracy consistently exhibit weaker data masking performance, and vice versa. This observation underscores the distinct capabilities required for these two objectives and highlights the urgent need for architectural approaches that jointly optimize operational utility and data confidentiality. Ultimately, \Design{} provides a unified framework for the rigorous evaluation of edge-deployed LLMs in real-world, security-critical environments.

\section*{Acknowledgements}
This work has been funded in part by NSF, with award numbers \#2112665, \#2112167, \#2003279, \#2120019, \#2211386, \#2052809, \#1911095 and in part by PRISM and CoCoSys, centers in JUMP 2.0, an SRC program sponsored by DARPA.

\bibliographystyle{IEEEtran}
\bibliography{biblio.bib}

\end{document}